\documentclass[aps,prl,amsmath,amssymb,floatfix,twocolumn,amsmath,superscriptaddress,twocolumn,nofootinbib,tighten,letterpaper]{revtex4-2}
\usepackage[colorlinks,linkcolor=blue,citecolor=blue,urlcolor=blue]{hyperref}
\usepackage{multirow}
\usepackage{subfigure}
\usepackage{color}
\usepackage{mathrsfs}
\usepackage{hyperref}
\usepackage[normalem]{ulem}
\usepackage{bm}

\usepackage{amssymb}
\usepackage{amsmath}
\renewcommand\vec[1]{\ensuremath\boldsymbol{#1}}

\usepackage{tabularray}

\usepackage{array} 
\newcolumntype{P}[1]{>{\centering\arraybackslash}p{#1}}

\definecolor{RowColor}{rgb}{0.88,1,0.9}

\usepackage{amsfonts, relsize, color, mathtools, physics}
\usepackage{graphics}
\usepackage{graphicx}
\usepackage{subfigure}
\usepackage{color}
\usepackage{comment}

\begin{document}

\title{Hyperbolic lattices with mass disorder: Phases and phase transitions}

\author{Sheersh Sen}\thanks{These two authors contributed equally to this work}
\affiliation{Centre for Condensed Matter Theory, Department of Physics, Indian Institute of Science, Bengaluru 560012, India}

\author{Christopher A.\ Leong}\thanks{These two authors contributed equally to this work}
\affiliation{Department of Physics, Lehigh University, Bethlehem, Pennsylvania 18015, USA}

\author{Bitan Roy}
\affiliation{Department of Physics, Lehigh University, Bethlehem, Pennsylvania 18015, USA}

\date{\today}

\begin{abstract}
Nearest-neighbor (NN) tight-binding models (TBMs) on plaquette-centered $\{ 10,3\}$ (Schl\"afli symbol), $\{ 8,3\}$, and $\{ 8,4\}$ hyperbolic lattices on a Poincar\'e disk with open boundary conditions display a vanishing,  a finite, and a diverging density of states (DOS) near zero energy (band center), respectively, yielding a Dirac liquid, a Fermi liquid, and a flat band. Emergent bipartite nature of these lattices within the framework of NN-TBMs, allows us to scrutinize the impact of mass disorder on the electronic states and DOS therein. From extensive numerical calculations of the average and typical DOS using the kernel polynomial method, we show that hyperbolic Dirac liquid remains stable against weak mass disorder, undergoing a semimetal-to-metal quantum phase transition at moderate disorder, followed by an Anderson metal-to-insulator transition at even stronger disorder. The remaining two systems display only the latter transition. Critical exponents near all these transitions are found to be close to the ones mediated by on-site potential disorder.     
\end{abstract}

\maketitle

\emph{Introduction}.~The nearest-neighbor (NN) tight-binding model (TBM) is a canonical description for free fermions in quantum materials~\cite{AshcroftMermin}. On hyperbolic lattices (HLs), obtained from periodic tessellations of regular polygons of $p$ arms on a curved space with a constant negative curvature, each vertex of which is connected to $q$ equidistant NN vertices, such a model displays a variety of scalings for the density of states (DOS)~\cite{HL:1, HL:2, HL:3, HL:4, HL:5, HL:6, HL:7, HL:8, HL:9, HL:10, HL:11, HL:12, HL:13, HL:14}, a quantity governing their thermodynamic responses like specific heat and compressibility. Due to infinitely many possible realizations, resulting from an inequality $(p-2)(q-2)>4$, HLs are identified by the Schl\"afli symbol $\{ p,q\}$. Especially, when $p$ is an even integer, the NN-TBM endows HLs with an emergent sublattice structure in which the NN hopping is finite only between the sites from complementary $A$ and $B$ sublattices. However, such a sublattice labeling is arbitrary, enriching bipartite HLs with an Isinglike sublattice exchange symmetry. The system then preserves the time-reversal, particle-hole, and sublattice symmetries, and belongs to the symmetry class BDI~\cite{Ryu2010}.

\begin{figure}[b!]
    \includegraphics[width=0.95\linewidth]{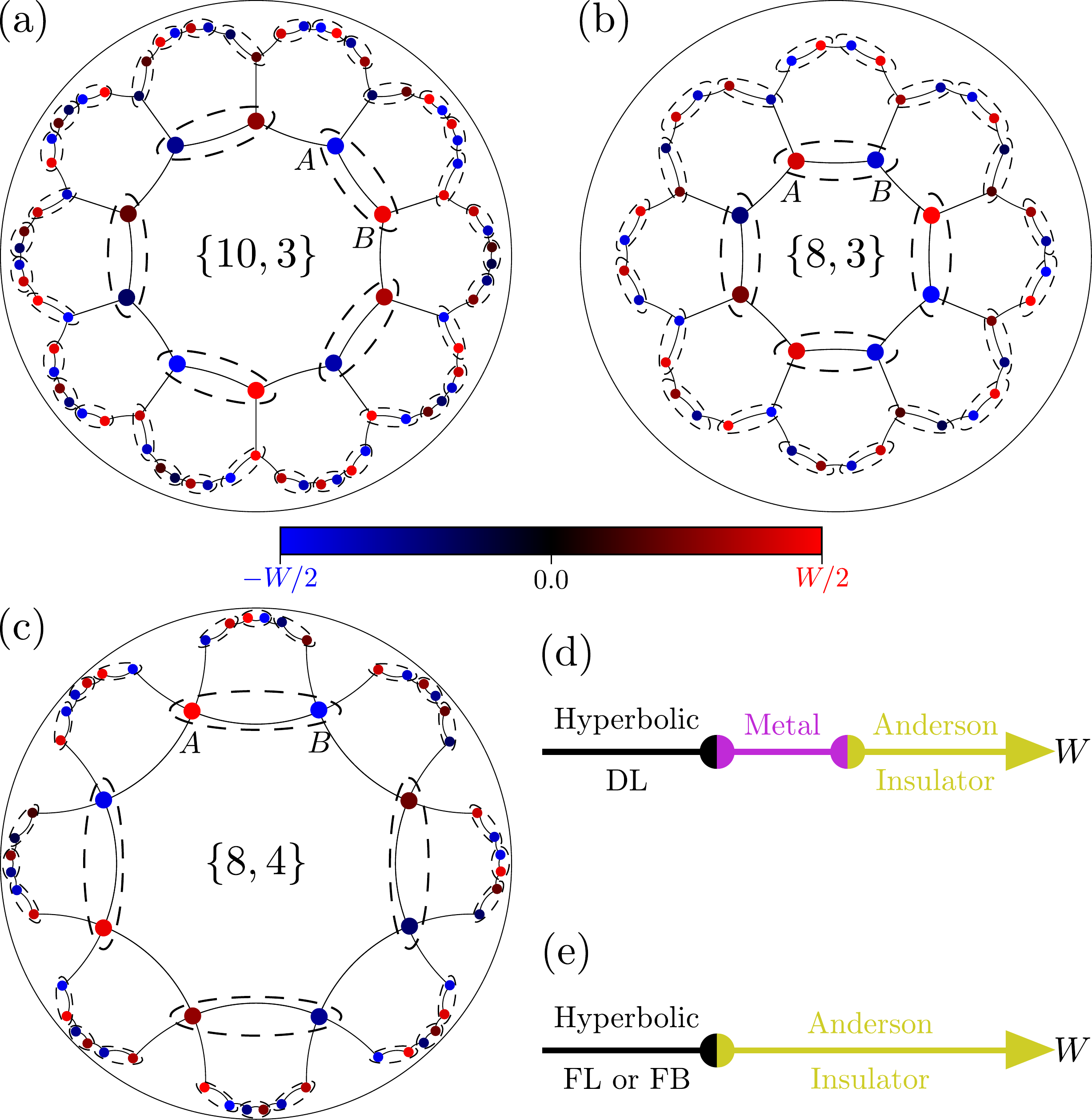}
    \caption{Schematic demonstration of mass disorder, represented by random on-site potential ($W$) of opposite signs (see color bar) on two sublattices ($A$ and $B$) within each unit cell (dashed closed loops) on plaquette-centered (a) $\{10,3\}$, (b) $\{8,3\}$, and (c) $\{8,4\}$ hyperbolic lattices on the Poincar\'e disk (black circle), respectively supporting a Dirac liquid (DL), a Fermi liquid (FL), and a flat band (FB) on a curved space with a constant negative curvature. The values of $W$ in different unit cells are independent and random. Schematic global phase diagrams in (d) a DL and (e) a FL and a FB in the presence of mass disorder with circles showing the quantum critical points between the adjacent phases.
    }~\label{fig:1_SchematicDiagrams}
\end{figure}

In particular, when plaquette-centered $\{ 10,3 \}$, $\{ 8,3 \}$, and $\{ 8,4 \}$ HLs are embedded on a Poincar\'e disk [Fig.~\ref{fig:1_SchematicDiagrams}] with \emph{open boundary conditions} (imposed throughout this work), the NN-TBM displays a power-law vanishing, a finite, and a diverging DOS near the band center, respectively, and are accordingly named a Dirac liquid (DL), a Fermi liquid (FL), and a flat band (FB)~\cite{HL:10, HL:12}. Therefore, a simple NN-TBM gives us an unprecedented opportunity to study electronic interactions~\cite{HL:10, HL:11, HL:12} and disorder~\cite{disHL:1, disHL:2, disHL:3, disHL:4, disHL:5} on such HLs to register signatures of spatial curvature on various emergent phenomena, triggered by inelastic and elastic scatterings, respectively.

\begin{figure}[t!]
    \includegraphics[width=0.95\linewidth]{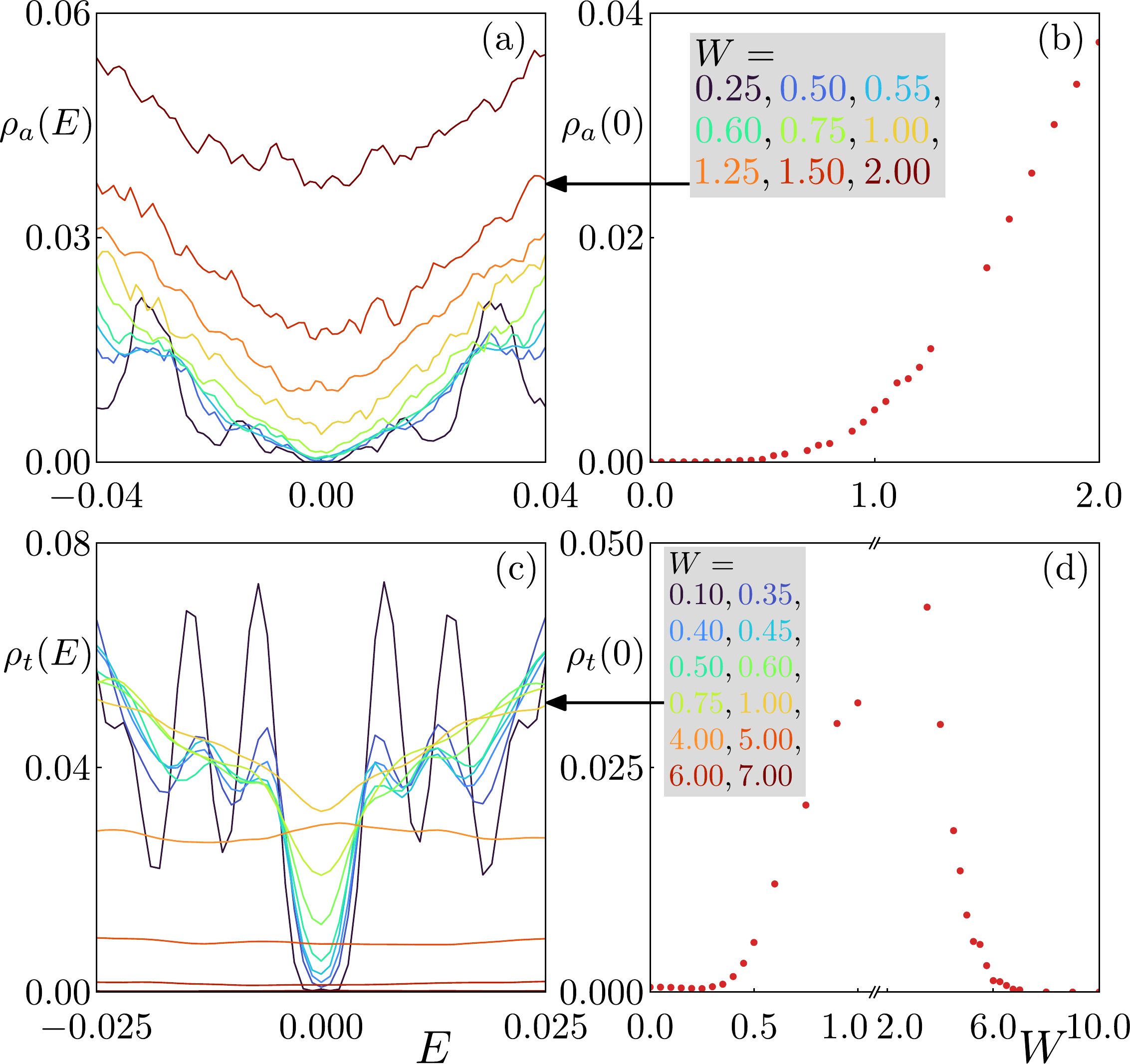}
    \caption{(a) ADOS $\rho_a(E)$ at energy $E$ with varying mass disorder ($W$) on a plaquette-centered $\{ 10,3 \}$ HL with open boundary conditions hosting a DL [Fig.~\ref{fig:1_SchematicDiagrams}(a)]. (b) $\rho_a(0)$ as a function of $W$, showing a semimetal-to-metal QPT at a critical disorder strength of $W_{c,1} \approx 0.55$. (c) TDOS $\rho_t(E)$ at energy $E$ with varying $W$ in the same system. (d) $\rho_t(0)$ with varying $W$ showing $W_{c,1} \approx 0.40$ and an AMIT at even stronger disorder, for which the critical disorder strength ($W_{c,2}$) depends on whether $\rho_t(0)$ shows a power-law (yielding $W_{c,2} \approx 7.00$) or an essential (yielding $W_{c,2} \approx 10.45$) singularity. ADOS and TDOS are computed using KPM after averaging over $10$ and $60$ independent disorder realizations, respectively, and with $6144$ Chebyshev moments. See Fig.~\ref{fig:1_SchematicDiagrams}(d). 
    }~\label{fig:2_p10q3DOS}
\end{figure}

\emph{Key results}.~The bipartite nature of $\{ 10,3 \}$, $\{ 8,3 \}$, and $\{ 8,4 \}$ HLs permits a simple realization of the mass disorder therein in the following way. On each unit cell a pair of random on-site potential $\pm V(\vec{r})$, otherwise drawn from a box distribution $[-W/2, W/2]$, is assigned arbitrarily to complementary sublattices, which varies randomly across the system among all the unit cells, as shown schematically in Figs.~\ref{fig:1_SchematicDiagrams}(a)-\ref{fig:1_SchematicDiagrams}(c). Employing the kernel polynomial method (KPM)~\cite{KPM:RMP}, we numerically compute both average density of sates (ADOS) and typical density of states (TDOS) at energy $E$, respectively denoted by $\rho_a(E)$ and $\rho_t(E)$, with varying strength of such mass disorder ($W$) and arrive at the following conclusions.

\begin{figure}[t!]
    \includegraphics[width=0.95\linewidth]{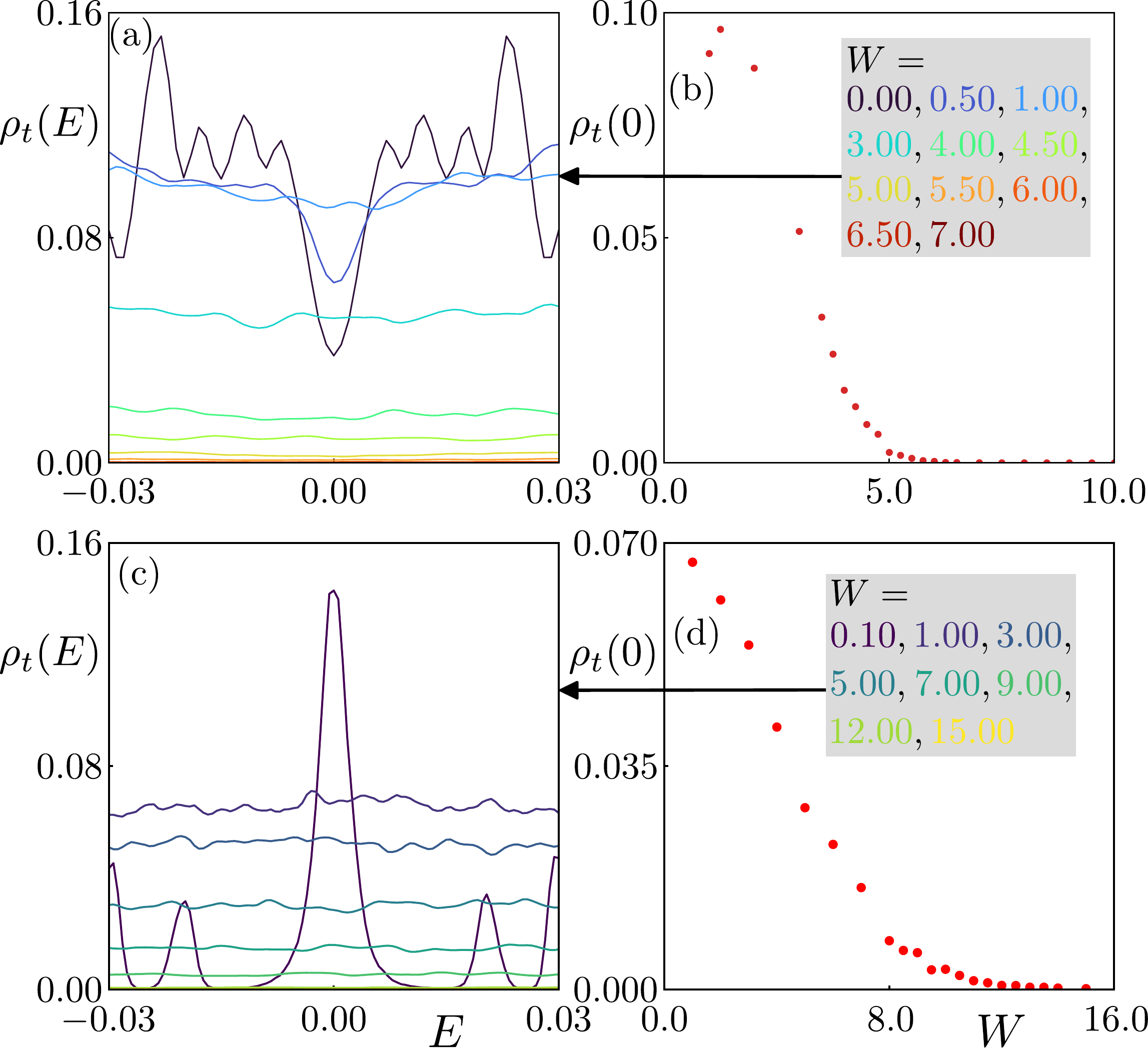}
    \caption{(a) TDOS $\rho_t(E)$ at energy $E$ as a function of mass disorder ($W$) in a plaquette-centered $\{8,3 \}$ HL with open boundary conditions hosting a FL [Fig.~\ref{fig:1_SchematicDiagrams}(b)]. (b) $\rho_t(0)$ with varying $W$, showing an AMIT for which the critical disorder strength ($W_{c,2}$) depends on whether $\rho_t(0)$ shows a power-law (yielding $W_{c,2} \approx 6.00$) or an essential (yielding $W_{c,2} \approx 8.25$) singularity. (c) and (d) are analogous to (a) and (b), respectively, but in a plaquette-centered $\{ 8,4 \}$ HL with open boundary conditions hosting a FB [Fig.~\ref{fig:1_SchematicDiagrams}(c)]. In (d), $W_{c,2} \approx 12.00$ ($W_{c,2} \approx 14.35$) if $\rho_t(0)$ shows a power-law (an essential) singularity near the AMIT. Results are obtained using the KPM with $8192$ ($16384$) Chebyshev moments on $\{ 8,3 \}$ ($\{ 8,4 \}$) HL and after averaging over $60$ independent disorder realizations.    
    }~\label{fig:3_NonDiracDOS}
\end{figure}

\begin{figure*}[t!]
    \includegraphics[width=0.975\linewidth]{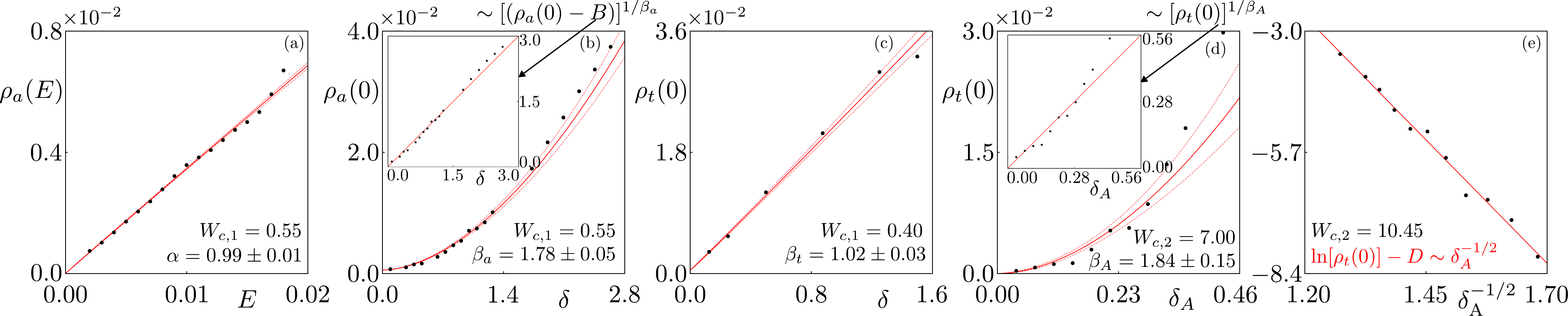}
    \caption{Scaling analysis on $\{10,3 \}$ HL subject to mass disorder. (a) ADOS exponent ($\alpha$), obtained from $\rho(E) \sim |E|^\alpha$ at the semimetal-to-metal quantum critical point, located at $W_{c,1}=0.55$, yielding $\alpha=0.99 \pm 0.01$. Order parameter exponent associated with (b) ADOS (with $W_{c,1}=0.55$) and (c) TDOS (with $W_{c,1}=0.40$) at $E=0$ in the metallic phase by comparing $\rho_a(0)-B \sim |\delta|^{\beta_a}$ with $B=5.65 \times10^{-4}$ as an offset and $\rho_t(0)-C \sim |\delta|^{\beta_t}$ with $C=4.84\times10^{-4}$ as an offset, respectively, where $\delta=[W-W_{c,1}]/W_{c,1}$, yielding $\beta_a=1.78 \pm 0.05$ [also confirmed from the linear scaling of $\left[ \rho_a(0)-B \right]^{1/\beta_a}$ with $\delta$ for $\beta_a=1.78$ (inset)] and $\beta_t=1.02 \pm 0.03$. (d) Order parameter exponent ($\beta_A$) across the AMIT assuming a power-law singularity of TDOS at $E=0$ with the scaling form $\rho_t(0) \sim \delta^{\beta_A}_A$, where $\delta_A=[W_{c,2}-W]/W_{c,2}$ with $W_{c,2}=7.00$, yielding $\beta_A=1.84 \pm 0.15$, also confirmed from the linear fit of $\left[ \rho_t(0) \right]^{1/\beta_A}$ with $\delta_A$ (inset). (e) Essential singularity of $\rho_t(0)$ near AMIT gives a good linear fit between $\ln\rho_t(0)-D$ and $\delta^{-1/2}_A$ for $W_{c,2}=10.45$ and $D=10.65$. 
    }~\label{fig:4_ScalingAnalysis103}
\end{figure*}

\begin{figure}[t!]
    \includegraphics[width=0.95\linewidth]{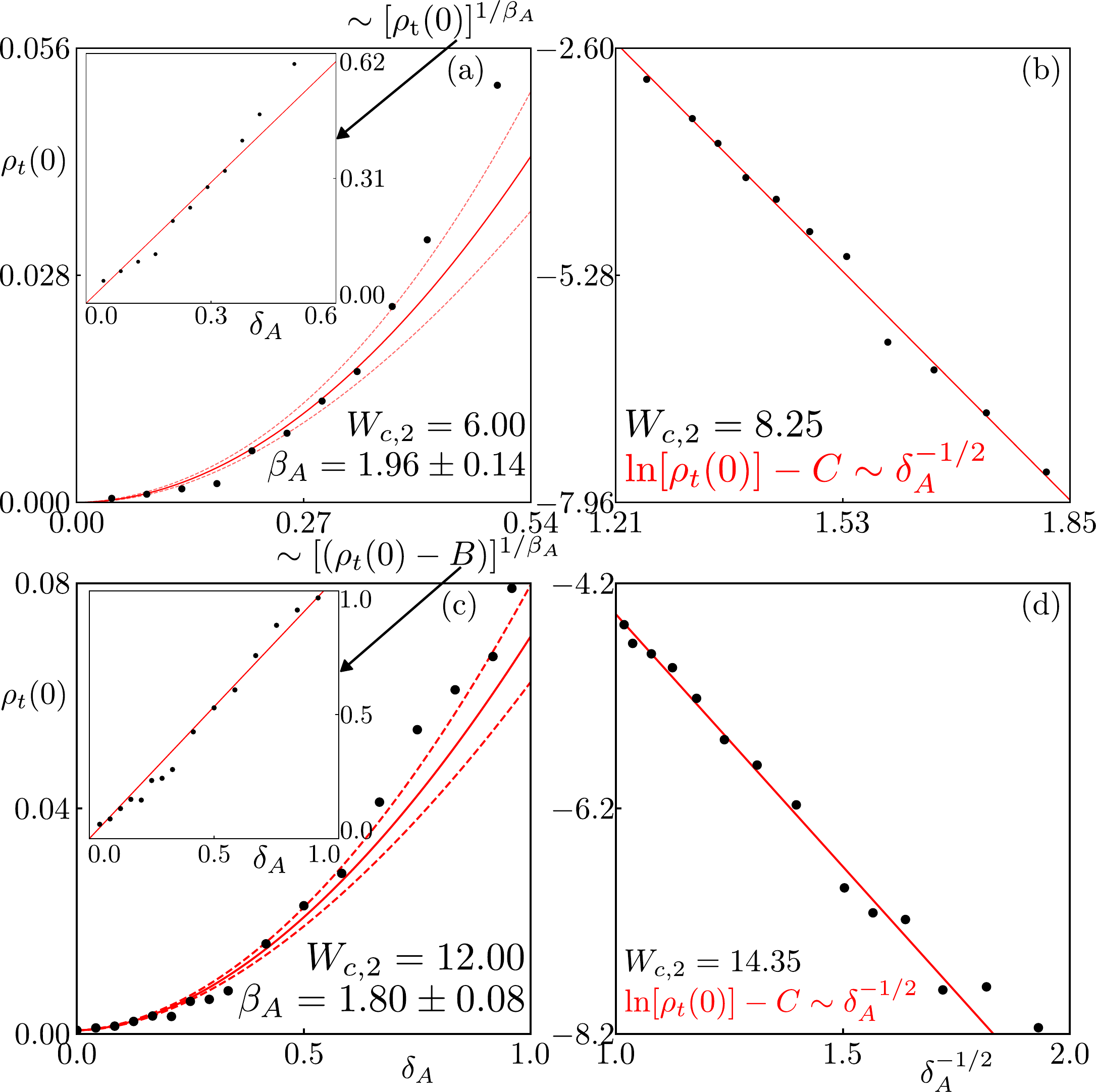}
    \caption{Scaling analysis on mass disordered $\{8,3\}$ [(a),(b)] and $\{8,4\}$ [(c),(d)] HLs near AMIT. (a) Order parameter exponent ($\beta_A$) assuming a power-law scaling of TDOS at $E=0$ form $\rho_t(0)-B \sim \delta^{\beta_A}_A$, where $\delta_A=[W_{c,2}-W]/W_{c,2}$ with $W_{c,2}=6.00$ and $B=0$, yielding $\beta_A=1.96 \pm 0.14$, also confirmed from the linear fit of $\left[ \rho_t(0)\right]^{1/\beta_A}$ with $\delta_A$ (inset). (b) Essential singularity of $\rho_t(0)$ gives a good linear fit between $\ln\rho_t(0)-C$ and $\delta^{-1/2}_A$ for $W_{c,2}=8.25$ and $C=8.19$. Panels (c) and (d) are analogous to (a) and (b), respectively, but with $W_{c,2}=12.00$, $\beta_A=1.80 \pm 0.08$, and $B=6.49 \times 10^{-4}$ in (c), and $W_{c,2}=14.35$ and $C=2.03$ for (d).
    }~\label{fig:5_ScalingAnalysis8384}
\end{figure}

\begin{figure}[t!]
    \includegraphics[width=0.95\linewidth]{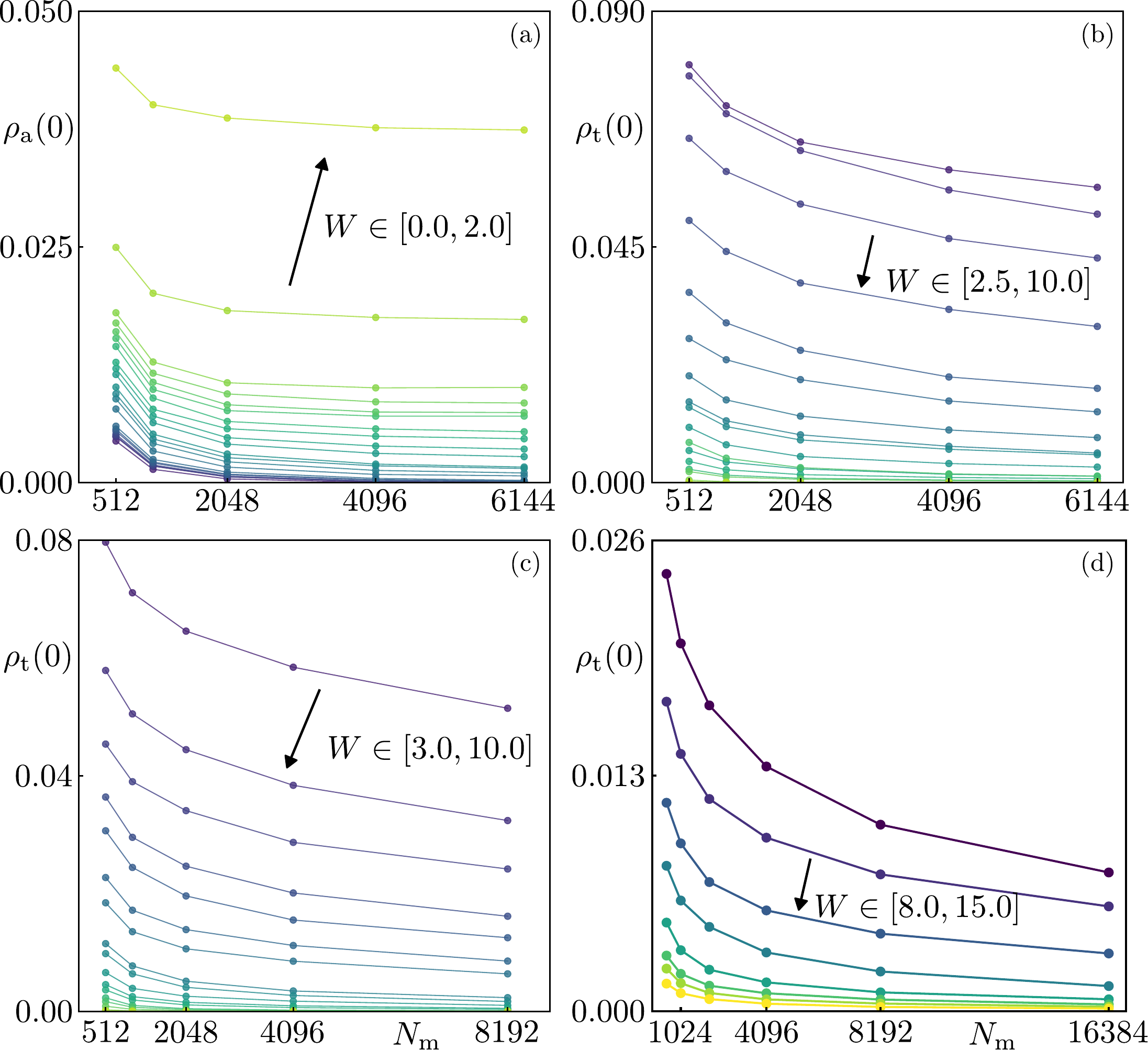}
    \caption{(a) Convergence of ADOS at zero energy $\rho_{\rm a}(0)$ as a function of the number of moments ($N_m$) used in reconstruction of the Chebyshev polynomial expansion for the $\{10, 3\}$ HL, for various strengths of the mass disorder $W$ (see colored curves). Analogous convergence plots for TDOS at zero energy $\rho_{\rm t}(0)$ for (b) $\{10, 3\}$, (c) $\{8, 3\}$, and (d) $\{8, 4\}$ HLs.}~\label{fig:6_MomentsConvergence}
\end{figure}

A hyperbolic DL remains stable for sufficiently weak mass disorder, where both $\rho_a(0)$ and $\rho_t(0)$ are vanishingly small (namely, $\lesssim 10^{-3}$). Beyond a moderate $W$ (namely, $W_{c,1}$) such a semimetallic system becomes a diffusive metal via a quantum phase transition (QPT), where both $\rho_a(0)$ and $\rho_t(0)$ become finite. As the strength of mass disorder is increased further, the system encounters the Anderson metal-to-insulator transition (AMIT) at even stronger $W$, denoted by $W_{c,2}$, where only $\rho_t(0)$ becomes zero. See Fig.~\ref{fig:1_SchematicDiagrams}(d) for a schematic global phase diagram for dirty $\{ 10,3 \}$ HLs and Fig.~\ref{fig:2_p10q3DOS} for the variation of DOS. By contrast, mass disordered hyperbolic FL and FB only display an AMIT [Fig.~\ref{fig:3_NonDiracDOS}] with their unique global phase diagram schematically shown in Fig.~\ref{fig:1_SchematicDiagrams}(e). We also extract various critical exponents across the semimetal-to-metal QPT and AMIT in $\{ 10, 3\}$ system [Fig.~\ref{fig:4_ScalingAnalysis103}], and across the AMIT in $\{ 8,3 \}$ and $\{ 8,4 \}$ systems [Fig.~\ref{fig:5_ScalingAnalysis8384}], which are sufficiently close to the ones found in the presence of on-site potential disorder~\cite{disHL:4}, as both types of disorder take these systems to class AI. We ensure numerical convergence of ADOS and TDOS with the number of Chebyshev moments [Fig.~\ref{fig:6_MomentsConvergence}].

\emph{Model}.~The NN-TBM on bipartite HLs in the presence of mass disorder reads as 
\begin{equation}
H= -t \sum_{\langle i,j \rangle} c^\dagger_i c_j + \left[ \; \sum_{i \in A} V(\vec{r}_i) c^\dagger_i c_i- \sum_{i \in B} V(\vec{r}_i) c^\dagger_i c_i\right].
\end{equation}
Here, $c^\dagger_i$ ($c_i$) is the fermionic creation (annihilation) operator on the $i$th site, $t$ is the real NN hopping amplitude (set to be unity), $\langle \cdots \rangle$ restricts the summation within the NN sites. To realize mass disorder, the value of the on-site potential $V(\vec{r}_i)$ on $A$ and $B$ sublattice sites within a unit cell are chosen to be of equal in magnitude but of opposite signs, which otherwise takes random values from a box distribution $[-W/2,W/2]$ over the unit cells across the system, as shown in Fig.~\ref{fig:1_SchematicDiagrams}(a)-\ref{fig:1_SchematicDiagrams}(c).

In an $N$-component spinor basis $\Psi^\top=(c_A, c_B)$, where $c_A$ ($c_B$) is an $N/2$-dimensional spinor constituted by the annihilation operators on the sites belonging to the $A$ ($B$) sublattice in an HL with $N$ sites, the Hermitian matrix associated with the NN-TBM in clean systems takes the form 
$\hat{h}_0 = \left( \begin{array}{cc}
{\boldsymbol 0} & {\boldsymbol t} \\
{\boldsymbol t}^\top & {\boldsymbol 0} 
\end{array}
\right)$.
Here, ${\boldsymbol 0}$ is an $N/2$-dimensional null matrix, ${\boldsymbol t}$ is an $N/2$-dimensional \emph{real} inter-sublattice hopping matrix, and $\top$ represents transposition. Notice that $\hat{h}_0$ preserves (a) the time reversal symmetry, generated by ${\mathcal T}= {\mathbb I}_N \; {\mathcal K}$, where ${\mathbb I}_N$ is an $N$-dimensional identity matrix and ${\mathcal K}$ is the complex conjugation operator, such that $[\hat{h}_0, {\mathcal T}]=0$ with ${\mathcal T}^2=+1$ (orthogonal symmetry class), (b) an antiunitary particle-hole symmetry, generated by ${\mathcal C}= \Sigma_z {\mathcal K}$, where $\Sigma_z={\rm diag}. ({\mathbb I}_{N/2}, -{\mathbb I}_{N/2})$ such that $\{ \hat{h}_0, {\mathcal C} \}=0$ and ${\mathcal C}^2=+1$, and (c) a unitary particle-hole or chiral or sublattice symmetry, generated by ${\mathcal S}=\Sigma_z$ such that $\{ \hat{h}_0, {\mathcal S} \}=0$. Therefore, $\hat{h}_0$ belongs to the \emph{chiral} orthogonal symmetry class BDI. With the addition of mass disorder, the system loses the particle-hole (${\mathcal C}$) and sublattice (${\mathcal S}$) symmetries, but continues to enjoy the ${\mathcal T}$ symmetry, thereby entering the \emph{Wigner-Dyson} orthogonal class AI~\cite{Ryu2010}, which does not permit any AMIT in $d=2$ on a Euclidean lattice~\cite{AMIT:1, AMIT:2, AMIT:3}.

\emph{System}.~Throughout we consider plaquette-centered $\{ 10, 3 \}$, $\{ 8, 3 \}$, and $\{ 8, 4 \}$ HLs, tiled on a Poincar\'e disk with open boundary condition and total generation $n_{\rm tot}=7$, 9, and 7, respectively; each system containing more than $10^5$ sites. Here, the central $p$-gon constitutes the first generation ($n=1$) and each successive layer of plaquettes constitutes its progressively next generation. In such a large system, the ratio of the number of bulk sites (characterized by the coordination number $q$) to the boundary sites (with coordination number $< q$) appraoches its values in the thermodynamic limit, which are $0.414$, $0.732$, and $0.207$ for $\{ 10, 3 \}$, $\{ 8, 3 \}$, and $\{ 8, 4 \}$ HLs, respectively. Thus our proposed scaling of the DOS in these HLs may depend on the boundary condition (open or periodic), but are independent of the system size.

\emph{DOS}.~We employ the KPM  to compute ADOS and TDOS in all these systems. The disorder-averaged, denoted by $\langle ... \rangle$, ADOS at energy $E$ is defined as 
\begin{equation}
    \rho_{a} (E) = \biggl< \frac{1}{N} \sum^{N}_{j=1} \; \delta \left( E-E_j \right) \: \biggr>.
\end{equation}
Even though, $\rho_{a} (E)$ is a self-averaging quantity, we still average it over 10 independent disorder realizations to minimize the residual statistical errors. While extracting $\rho_a(E)$, we compute $N_m$ chebyshev moments by taking stochastic trace over 12 unimodular random vectors. The local DOS at the $i$th site at energy $E$ is given by
\begin{equation}
\rho^{i}_{\rm loc} (E) = \sum^{N}_{j=1} |\langle E_j|i\rangle|^2 \: \delta\left( E-E_j\right), 
\end{equation}
where $\ket{i}$ is the site-localized Wannier wavefunction, and $\ket{E_j}$ is the eigenstate with energy $E_j$. The TDOS over a $p$-gon from the $n$th generation of a HL is defined as 
\begin{equation}
\rho^{n}_{t} (E) = \exp\left( \frac{1}{p}\sum^{p}_{i=1} \bigl< \ln \rho^{i}_{\rm loc} (E) \bigr>  \right). 
\end{equation}
As $\rho^{n}_{t} (E)$ is not a self-averaging quantity, we average it over 60 independent disorder realizations. We compute TDOS in each generation of the HLs to scrutinize the role of the large number of boundary sites, showing that results do not qualitatively depend on $n$. Finally, we compute the \emph{average} TDOS, defined as $\rho_{t}(E)=\sum^{n_{t}}_{n=1} \rho^{n}_{t} (E)/n_{\rm tot}$. Throughout, the computation of Chebyshev moments is augmented by the Jackson kernel to minimize the Gibbs oscillations~\cite{KPM:RMP}.

\emph{Phase diagram}.~The global phase diagram of mass disordered HLs, featuring DL, FL, and FB in clean systems can be assessed from the scaling of ADOS and TDOS at zero energy. On $\{ 10,3 \}$ HL, $\rho_a(0)<10^{-3}$ [$\rho_t(0)<10^{-2}$], thus remains pinned close to zero, up to $W=0.55$ [$W=0.40$], beyond which it becomes finite, thereby marking the critical disorder for the semimetal-to-metal QPT $W_{c,1}$ (Fig.~\ref{fig:2_p10q3DOS})~\cite{SM:Massdisorder}. Slightly different values of $W_{c,1}$ obtained from ADOS and TDOS stems from the stability of KPM, which strictly sets in at disorder strength $\gtrsim  1$. Unfortunately, $W_{c,1}$ is relatively small for the best stability of KPM, otherwise expected to give \emph{identical} $W_{c,1}$ from the scaling of $\rho_a(0)$ and $\rho_t(0)$~\cite{disHL:4}. Still the difference in values of $W_{c,1}$ is sufficiently small.

The stability of DLs against sufficiently weak mass disorder and their transition to a metal, are \emph{possibly} rooted in the relativistic nature of quasiparticles living on a negative curved space with a constant curvature $a$, for which the DOS scales as $\rho(E) \sim E \tanh(a E) \to |E|^2$ at small energies~\cite{HL:2, disHL:4}. Then the solution of quasiparticle lifetime ($\tau_{\rm qp}$), obtained within the self-consistent Born approximation~\cite{Book:SCBA}, leading to $W \int^{E_\Lambda}_0 \rho(E)/[E^2+\tau^{-2}_{\rm qp}] dE=1$ for $\hbar=1$, where $E_\Lambda$ is an ultraviolet energy cutoff up to which $\rho(E) \sim |E|^2$, gives a finite $\tau^{-1}_{\rm qp} \propto \rho(0)$ only beyond a critical disorder strength $W_\star=\big[ \int^{E_\Lambda}_0 \rho(E)/[E^2] dE \big]^{-1}$. Then, taking  $E_\Lambda \to \infty$, we obtain $\tau^{-1}_{\rm qp}=(2/\pi) \bar{\delta}$, where $\bar{\delta}=W^{-1}_\star-W^{-1}$. Such a solution of $\tau^{-1}_{\rm qp}$ in turn endorses the stability of DL on HLs for sufficiently weak disorder (with $\tau^{-1}_{\rm qp}=0$ for $\bar{\delta}<0$) and its QPT into a metal (with a finite $\tau^{-1}_{\rm qp}$ for $\bar{\delta}>0$) at a moderate disorder ($W_\star$)~\cite {disHL:4}.

As the strength of disorder is increased further there exists an AMIT at even stronger disorder ($W_{c,2}$) solely due to the underlying spatial curvature~\cite{disHL:1, disHL:2, disHL:3, disHL:4, disHL:5}. Recall that in two-dimensional (2D) Euclidean lattices, there is no AMIT in class AI~\cite{AMIT:1, AMIT:2, AMIT:3}. A finite spatial curvature provides an infrared cutoff to the return probability of a particle to its original position to interfere and localize. As a result, on 2D HLs a metallic phase can be stable for moderate disorder, which ultimately undergoes an AMIT at sufficiently strong disorder, across which only $\rho_t(0)$ vanishes. Such an outcome is independent of whether the clean HL fosters a DL or FL or FB. Results are shown in Figs.~\ref{fig:2_p10q3DOS}(c) and~\ref{fig:2_p10q3DOS}(d) for $\{ 10,3 \}$ HL, in Figs.~\ref{fig:3_NonDiracDOS}(a) and~\ref{fig:3_NonDiracDOS}(b) for $\{ 8,3 \}$ HL, and in Figs.~\ref{fig:3_NonDiracDOS}(c) and~\ref{fig:3_NonDiracDOS}(d) for $\{ 8,4 \}$ HL. Qualitatively similar results for TDOS are found when computed on a single site from each generation of the HL~\cite{SM:Massdisorder}. We note that $\{ 8,4 \}$ HL continues to support a prominent DOS peak near $E=0$ (residue of the clean system FB) up to a moderate disorder [Fig.~\ref{fig:3_NonDiracDOS}(c)].

These findings allows us to promote the global phase diagram of electronic fluids realized on bipartite HLs. Namely, dirty hyperbolic Dirac systems feature three distinct phases; a semimetal, a metal, and an Anderson insulator, with two quantum critical points separating them [Fig.~\ref{fig:1_SchematicDiagrams}(d)]; mimicking closely in that respect the global phase diagram of three-dimensional disordered Euclidean Dirac-Weyl semimetals~\cite{disDiracWeyl:1, disDiracWeyl:2, disDiracWeyl:3, disDiracWeyl:4, disDiracWeyl:5, disDiracWeyl:6, disDiracWeyl:7, disDiracWeyl:8, disDiracWeyl:9}. By contrast, disordered hyperbolic FL and FB systems support only two stable phases of matter; a metal and an Anderson insulator, separated by a single quantum critical point [Fig.~\ref{fig:1_SchematicDiagrams}(e)]; closely resembling the phase diagram of three-dimensional dirty Euclidean metals~\cite{AMIT:1, AMIT:2, AMIT:3}.

\emph{Scaling}.~Near the semimetal-to-metal QPT in DL, we define the DOS exponent ($\alpha$) from $\rho_a(E) \sim |E|^{\alpha}$, yielding $\alpha=0.99 \pm 0.01$ [Fig.~\ref{fig:1_SchematicDiagrams}(a)]. The order parameter exponents $\beta_a$ and $\beta_t$ defined from the scaling of ADOS and TDOS at $E=0$, respectively, as $\rho_a(0) \sim \delta^{\beta_a}$ and $\rho_t(0) \sim \delta^{\beta_t}$ with $\delta=(W-W_{c,1})/W_{c,1}$, are found to be $\beta_a= 1.78 \pm 0.05$ for $W_{c,1}=0.55$ [Fig.~\ref{fig:4_ScalingAnalysis103}(b)] and $\beta_t= 1.02 \pm 0.03$ for $W_{c,1}=0.40$ [Fig.~\ref{fig:4_ScalingAnalysis103}(c)]. Distinct values of $\beta_a$ and $\beta_t$ suggest that wavefunctions near such a QPT are possibly multifractal in nature~\cite{MirlinEvers:RMP}. Power-law scaling of $\rho_t(0)$ in $\{ 10,3 \}$ HL near the AMIT yields the corresponding order parameter exponent $\beta_A=1.84 \pm 0.15$, defined as $\rho_t (0) \sim \delta^{\beta_A}_A$, where $\delta_A=(W_{c,2}-W)/W_{c,2}$ and $W_{c,2}=7.00$ [Fig.~\ref{fig:4_ScalingAnalysis103}(d)]. However, due to the presence of a large number sites at the boundary of HLs, which makes their effective dimensionality \emph{infinite}, an \emph{essential singularity} in TDOS of the form $\rho_t(0) \sim \exp[-\lambda/\sqrt{W_{c,2}-W}]$, where $\lambda$ is a constant, cannot be ruled out~\cite{essential:review}. And we find good agreement with such a scaling form on $\{ 10,3\}$ HLs, but, with a larger value of $W_{c,2}=10.45$, which is expected due to the exponential decay of $\rho_t(0)$ near AMIT [Fig.~\ref{fig:4_ScalingAnalysis103}(e)].

Similar scaling analyses near the AMIT yield $\beta_A=1.96 \pm 0.14$ with $W_{c,2}=6.00$ on $\{ 8,3 \}$ HL [Fig.~\ref{fig:5_ScalingAnalysis8384}(a)] and $\beta_A=1.80 \pm 0.08$ with $W_{c,2}=12.00$ on $\{ 8,4 \}$ HL [Fig.~\ref{fig:5_ScalingAnalysis8384}(c)]. Therefore, the values of $\beta_A$ on all three HLs are almost equal (within numerical errorbars). Finally, notice that even on $\{ 8,3 \}$ and $\{ 8,4 \}$ HLs we find good agreements with the essential singular form of the TDOS at $E=0$, as shown in Figs.~\ref{fig:5_ScalingAnalysis8384}(b) and~\ref{fig:5_ScalingAnalysis8384}(d), respectively, and expectedly with larger values of $W_{c,2}=8.25$ and $14.35$. Throughout, we ensure the convergence of ADOS and TDOS with respect to the number of Chebyshev moments in KPM-based calculations [Fig.~\ref{fig:6_MomentsConvergence}].

\emph{Summary and discussions}.~To summarize, from extensive numerical calculations of ADOS and TDOS using the KPM, we showcase the impacts of mass disorder on the electronic states of a family of bipartite HLs that in the clean limit harbor a DL, a FL, and a FB, respectively identified from a power-law vanishing, a constant, and a diverging DOS. While a DL supports three distinct stable phases or matter, namely a semimetal, a metal, and an Anderson insulator, the FL and FB systems sustains only the latter two phases of matter. The critical exponents across the semimetal-to-metal transition and AMIT, triggered by mass disorder are sufficiently close to the ones, when such QPTs are driven by on-site potential disorder~\cite{disHL:4}, as both types of disorder take these systems to class AI. We note that the across the AMIT, TDOS may display an essential singularity, instead of a power-law scaling, which requires further investigations.

Any classical metamaterial on which HLs can be engineered and that fosters wave dynamics~\cite{AndTransPhotonics:1}, constitutes a natural testbed for our theoretical predictions, among which photonic lattices are the most promising~\cite{AndTransPhotonics:2}. On such a platform HLs have already been designed~\cite{photonichyperboliclattices:1}. Due to the transverse nature of photon propagation and the absence of AMIT on Euclidean lattices, three-dimensional Euclidean disordered photonic lattices only show Anderson localization of light~\cite{AndTransPhotonics:3, AndTransPhotonics:4, AndTransPhotonics:5}. By contrast, our study shows that on dirty photonic HLs with open boundary conditions can display both semimetal-to-metal QPT (on $\{ 10,3 \}$ HL) and AMIT (on $\{ 10,3 \}$, $\{ 8,3 \}$, and $\{ 8,4 \}$ HLs) when the refractive index of the hosting material is randomly varied in the system following the intra-unit cell staggered pattern shown in Fig.~\ref{fig:1_SchematicDiagrams}(a)-\ref{fig:1_SchematicDiagrams}(c).

\emph{Acknowledgments}.~S.S.\ was supported by the INSPIRE-SHE program of DST, India. C.A.L.\ and B.R.\ were supported by NSF CAREER Grant No.\ DMR-2238679. Portions of this research were conducted on Lehigh University's Research Computing infrastructure partially supported by NSF Award No.~2019035. B.R.\ is indebted to Vladimir Dobrosavljevi{\ifmmode \acute{c}\else \'{c}\fi{}} for drawing attention to the essential singularity near AMIT~\cite{YongtaiLi:HL}.  

\emph{Data availability}.~Numerical codes and data used and generated in this work are available in Ref.~\cite{datacode:mass}.

\end{document}